\documentclass[]{aa}
\usepackage{graphics}
\usepackage{psfig}

\newcommand{\cmcub}{~cm$^{-3}$}

\newcommand{\ergsa}{~erg~s$^{-1}$~cm$^{-2}$~\AA$^{-1}$}

\newcommand{\teff}{$T_{\rm eff}$}

\newcommand{\Zs}{~Z$_{\odot}$}
\newcommand{\Ms}{~M$_{\odot}$}

\newcommand{\msun}{\ifmmode M_{\odot} \else M$_{\odot}$\fi}
\newcommand{\rsun}{\ifmmode R_{\odot} \else R$_{\odot}$\fi}
\newcommand{\lsun}{\ifmmode L_{\odot} \else L$_{\odot}$\fi}
\newcommand{\zsun}{\ifmmode Z_{\odot} \else Z$_{\odot}$\fi}

\newcommand{\qh}{$Q({\rm{H^{0}}})$}
\newcommand{\qhe}{$Q({\rm{He^{0}}})$}
\newcommand{\qhep}{$Q({\rm{He^{+}}})$}

\newcommand{\Teff}{T$_{eff}$}

\newcommand{\Ha}{H$\alpha$}
\newcommand{\Hb}{\ifmmode {\rm H}\beta \else H$\beta$\fi}
\newcommand{\hi}{H~{\sc i}}
\newcommand{\hii}{H~{\sc ii}}
\newcommand{\heii}{He~{\sc ii}}
\newcommand{\Heii}{He~{\sc ii} $\lambda$4686}
\newcommand{\Ciii}{C~{\sc iii}] $\lambda$1909}

\newcommand{\Nii}{[N~{\sc ii}] $\lambda$6584}

\newcommand{\Oi}{[O~{\sc i}] $\lambda$6300}
\newcommand{\oi}{[O~{\sc i}]}
\newcommand{\Oii}{[O~{\sc ii}] $\lambda$3727}
\newcommand{\oii}{[O~{\sc ii}]}
\newcommand{\Oiii}{[O~{\sc iii}] $\lambda$5007}
\newcommand{\Oiiit}{[O~{\sc iii}] $\lambda$4363}
\newcommand{\oiii}{[O~{\sc iii}]}
\newcommand{\Neiii}{[Ne~{\sc iii}] $\lambda$3869}

\newcommand{\Sii}{[S~{\sc ii}] $\lambda$6725}
\newcommand{\sii}{[S~{\sc ii}]}
\newcommand{\Siii}{[S~{\sc iii}] $\lambda$9532}
\newcommand{\Siiit}{[S~{\sc iii}] $\lambda$6312}
\newcommand{\siii}{[S~{\sc iii}]}
\newcommand{\Ariii}{[Ar~{\sc iii}] $\lambda$7135}

\newcommand{\rOiii}{[O~{\sc iii}] $\lambda$4363/5007}
\newcommand{\rNii}{[N~{\sc ii}] $\lambda$5755/6584}
\newcommand{\rSii}{[S~{\sc ii}] $\lambda$6731/6717}
\newcommand{\rAriv}{[Ar~{\sc iv}] $\lambda$4741/4713}

\newcommand{\Ho}{H$^{0}$}

\newcommand{\Hep}{He$^{+}$}
\newcommand{\Hepp}{He$^{++}$}

\newcommand{\Np}{N$^{+}$}

\newcommand{\Op}{O$^{+}$}
\newcommand{\Opp}{O$^{++}$}
\newcommand{\Oppp}{O$^{+++}$}

\newcommand{\Sp}{S$^{+}$}

\newcommand{\Arppp}{Ar$^{+++}$}

\begin{document}

   \thesaurus{11     % A&A Section 11: Galaxies
              (11.01.1; %Galaxies: abundances,
               11.09.1; %Galaxies: individual
               11.09.4; % Galaxies: ISM,
               11.19.3; %Galaxies: starburst,
               08.05.1; %Stars: early-type,
               08.23.2)} %Stars: Wolf-Rayet,

   \title{What heats the bright \hii\ regions in I~Zw~18?}

   \author{G. Stasi\'{n}ska
\inst{1},  D. Schaerer \inst{2}}

\institute{ DAEC, Observatoire de Paris-Meudon, F-92195 Meudon Cedex,
 France\\email: 
grazyna@obspm.fr\and Observatoire de Midi-Pyr\'en\'ees, Laboratoire 
d'Astrophysique, 14 Av. E. Belin, F-31400 Toulouse, \\ email: 
schaerer@obs-mip.fr}
\date{Received 17 march 1999 / Accepted 10 september 1999}
\titlerunning{I\,Zw\,18}
\authorrunning{Stasi\'{n}ska \& Schaerer}
\offprints{G.\ Stasi\'{n}ska}
\maketitle

%\onecolumn
\begin{abstract}
We have used the  radiation field from a starburst population 
synthesis model appropriate for the brightest \hii\ region of I~Zw~18 
to perform a photoionization model analysis of this object. 
We have investigated whether, with the observed nebular density 
distribution as revealed by the {\em HST} images and a total stellar radiation
compatible with the observed UV flux,  
one could reproduce the constraints represented by the 
observed ionization structure, the \Heii/\Hb\ ratio, the \Ha\ flux
and the electron temperature indicated by the \rOiii\ ratio.  

We have found that, even taking into account strong deviations from 
the adopted 
spectral energy distribution of the ionizing radiation and the effect 
of additional X-rays, the photoionization models yield too low a \rOiii\ 
ratio by about 30\%. This discrepancy is significant and poses an interesting 
 problem, which cannot be solved by expected inaccuracies in the 
atomic data. The missing energy may be of the same order of magnitude 
as the one provided by the stellar photons 
or lower, depending on the way it acts on the \Oiiit\ line. 

Elemental abundance determinations in I~Zw~18 are affected by 
this problem. Before the nature of the missing heating source and its 
interactions with the nebular gas are better understood it is, however,
not possible to estimate the typical uncertainties by which 
standard empirical methods are affected.

Several side-products of our photoionization analysis of I~Zw~18 are 
the following: 
\begin{itemize}
\item  
We have been able to reproduce the intensity of the nebular \Heii\ 
using a stellar radiation field consistent with
the observed Wolf-Rayet features in I~Zw~18.  

\item 
We have shown that the bright NW \hii\ region in I~Zw~18  does not 
absorb all the ionizing photons from the central star cluster, and 
that about half of them are available to ionize an extended diffuse 
envelope.

\item
The \oi\ emission  can 
easily be accounted for by condensations or intervening filaments
on the line of sight. 

\item
The intrinsic \Ha/\Hb\ ratio is enhanced by collisional
 excitation to a value of 3. Previous estimates of the reddening 
 of I~Zw~18 are therefore slightly overestimated. 

\item
We give ionization correction factors 
appropriate for I~Zw~18 that can be used for more accurate abundance 
determinations in this object once the electron temperature problem 
is better understood. 
\end{itemize}

 \keywords{ 
          Galaxies: abundances --
          Galaxies: ISM --
          Galaxies: starburst --
          Galaxies: individual: I Zw 18 --
          Stars: early-type --
          Stars: Wolf-Rayet}.
\end{abstract}

\section{Introduction}

The blue compact dwarf emission-line galaxy I\,Zw\,18 is famous for 
being the most metal poor galaxy known so far. Its oxygen abundance 
is about 2\% the solar value, as first shown by Searle and Sargent 
(1972), and then confirmed by many studies (e.g. Lequeux et al. 1979, 
French 1980, Kinman \& Davidson 1981, Pagel et al. 1992, Legrand et 
al. 1997, Izotov et al. 1997b, V\'{\i}lchez 
\& Iglesias-P\'{a}ramo 1998). Because of this,  I\,Zw\,18 has played 
an essential role in the determination of the primordial helium mass 
fraction.  Also, due to its 
extreme properties, I\,Zw\,18 has been a choice target for studies of 
star formation history in blue compact galaxies (Dufour \& Hester 
1990, Hunter \& Thronson 1995, Dufour et al. 1996, De Mello et al. 
1998, Aloisi et al. 1999), of the elemental enrichment in dwarf 
galaxies (Kunth \& Sargent 1986, Kunth et al. 1995) and of the 
interplay between star formation and the interstellar medium (Martin 
1996, van Zee et al. 1998). An important clue is the distribution of 
the oxygen abundance inside the \hii\ regions (Skillman \& Kennicutt 
1993, V\'{\i}lchez \& Iglesias-P\'{a}ramo 1998) 
and in the neutral gas 
(Kunth et al. 1994, Pettini \& Lipman 1995, van Zee et al. 1998). 
Another clue is the carbon and nitrogen abundance (Garnett et al. 
1997, V\'{\i}lchez \& Iglesias-P\'{a}ramo 1998, Izotov \& Thuan 1999). 
On the whole, there is a 
general consent about an intense and recent 
burst of star formation in I\,Zw\,18  - which provides the ionizing 
photons - following previous star formation episodes. How exactly has 
the gas been enriched with metals during the course of the evolution 
of I\,Zw\,18  remains to be better understood. 

Much of our understanding (or speculations) on the chemical evolution 
of I\,Zw\,18  (and other galaxies in general) relies on the 
confidence 
placed in the chemical abundances derived from the lines emitted in 
the \hii\ regions. These are generally obtained using standard, 
empirical, 
methods which have been worked out years ago, and rely on the theory 
of line emission in 
photoionized gases. Photoionization models are, most of the time, 
used merely as a guide to evaluate 
the temperature of the low excitation regions once the characteristic 
temperature of the high excitation zones has been obtained through 
the \rOiii\ ratio. They 
also serve to provide formulae for estimating the correction factors 
for 
the unseen ionic species of a given element. 

Direct fitting of
the observed emission  line spectrum by tailored photoionization 
models provides more accurate
abundances only if all the relevant line ratios are perfectly
reproduced by the model
(which is rarely the case in model fitting history) and if the
model reproducing all the observational constraints is unique.

One virtue of model fitting, though, is that it permits to check 
whether the assumptions used in abundance determinations are correct 
for a given object. For example, there is the long standing debate 
whether so-called ``electron temperature fluctuations'' (see e.g. Mathis 1995,
Peimbert 1996, Stasi\'{n}ska 1998) are present 
in \hii\ regions to a 
sufficient level so as to significantly affect elemental abundance 
determinations. If a photoionization model is not able to reproduce 
all the temperature sensitive line ratios, the energy balance is not 
well understood, and one may question the 
validity of abundance determinations. Also, photoionization models 
are a potential tool (see e.g. Esteban et al. 1993,  Garc\'{\i}a-Vargas 
1996, Stasi\'nska \& Schaer\-er 1997, Crowther et al.\ 1999) 
to uncover the spectral distribution of 
the ionizing radiation field, thus providing information on the 
ionizing stars, their evolutionary status and the structure of their 
atmospheres. 

These two points are a strong motivation for a photoionization model 
analysis of I\,Zw\,18. There have already been a few such attempts in 
the past (Dufour et al. 1988, Campbell 1990, Stevenson et al. 1993). 
None of those models were, however, able to reproduce the \Heii\ 
line, 
known to exist in I\,Zw\,18 since the work of French (1980). The 
reason 
is that, in those models, the spectral distribution of the ionizing 
radiation was that of a single 
star whose radiation field was interpolated from a grid of
plane-parallel, LTE model atmospheres for massive stars. 
Recently,  Wolf-Rayet stars have been identified in I\,Zw\,18 through 
the characteristic bump they produce at 4650~\AA\ (Izotov et al. 1997a, 
Legrand et al. 1997). 
Spherically expanding non-LTE model atmospheres for hot Wolf-Rayet 
stars 
with sufficiently low wind densities (Schmutz et al. 1992) do predict 
an 
output of radiation above the \heii\ ionization edge, which might, at 
least 
qualitatively, provide a natural explanation for the narrow \Heii\ 
line 
observed in I\,Zw\,18. Schaerer (1996) has, for the first time, 
synthesized the broad (stellar) and narrow (nebular) \heii\ features 
in young starbursts using the Geneva stellar evolution tracks and 
appropriate stellar model atmospheres. He then extended his 
computations to the metallicity of I\,Zw\,18 (De Mello et al. 1998). 

In this paper, we use the emergent radiation field from the synthetic 
starburst model presented in De Mello et al. (1998) to construct 
photoionization models of I\,Zw\,18. One 
of the objectives is to see whether this more realistic ionizing 
radiation field permits, at the same time, to solve the electron 
temperature problem encountered in previous studies. Former 
photoionization models predicted too low a \rOiii\ ratio, unless 
specific geometries were adopted (Dufour et al. 1988, Campbell 1990), 
which 
later turned out to be incompatible with Hubble Space Telescope 
({\em HST}) images. 
The synthetic starburst model we use is based on spherically 
expanding non-LTE stellar atmosphere models for main sequence stars 
(Schaerer \& de Koter 1997) and for Wolf-Rayet stars (Schmutz et al.\
1992). These models have a greater heating power than the LTE model 
atmospheres of same effective temperature (see Fig.\ 3;
also Schaerer \& de Koter)

The progression of the paper is as follows. In Section 2, we  discuss 
in more detail the photoionization models proposed previously for 
I\,Zw\,18 and show in what respect they are not consistent with 
recent 
observations. In Section 3, we present our own model fitting 
methodology, including a description of the computational tools. In 
Section 4, we describe the models we have built for I\,Zw\,18, 
and discuss the effects of the assumptions involved in the computations. 
Our main results are summarized in Section 5.

\section{Previous photoionization models of I\,Zw\,18}

The first attempt to produce a photoionization model for I\,Zw\,18 is 
that of Dufour et al. (1988). Their observational constraints were 
provided by spectra obtained in an aperture of 2.5\arcsec $\times$
6\arcsec  of the NW region 
combined with IUE observations yielding 
essentially the \Ciii\ line. Using Shields's photoionization code 
NEBULA, they modelled the object as an ionization bounded sphere of 
constant density $n$ = 100\cmcub\ and adjustable volume filling 
factor 
$\epsilon$ so as to reproduce the observed \Oiii/\Oii\ ratio. The 
ionizing radiation was provided by a central source of radiation, 
represented by the LTE model atmospheres of Hummer \& Mihalas (1970), 
modified to take into account the low metallicity of I\,Zw\,18. 
Discarding the 
\heii\ problem, they obtained a model that was reasonably successful 
except that it had an \Opp\ temperature, T(\Opp), marginally smaller 
than observed 
(17200~K) compared to the value of 18100 (+1100, -1000)~K 
derived directly from their observed \rOiii\ = (2.79 $\pm$
0.35) $\times$ 10$^{-2}$ (the errors quoted being 2$\sigma$)
\footnote{A summary of various measurements of \rOiii\ and corresponding
electron temperatures is shown in Fig.\ 2.}.
This model was obtained for 
an effective temperature of 45000~K. These authors showed that, 
because of the 
dominant role played by Ly$\alpha$ cooling in I\,Zw\,18, it was 
impossible, for 
the adopted geometry, to produce a model with 
noticeably higher T(\Opp), by varying the free parameters at hand. 
Even increasing the effective temperature did not raise 
T(\Opp) appreciably, because then the ionization parameter had to be 
lowered in order to maintain \Oiii/\Oii\ at its observed level, and 
this resulted in a greater 
\Ho\ abundance, thus enhancing  Ly$\alpha$ excitation.  Dufour et al. 
then proposed a 
composite model, in which the \oiii\ line would be mainly produced 
around high temperature stars (\Teff\ $>$ 38000~K) and the \oii\ line 
would be mainly emitted around stars of lower \Teff\ ($<$ 37000~K). 
Alternatively, one could have, around a star of \Teff\ $<$ 45000~K, a 
high ionization component emitting most of the \oiii\ and a low 
ionization component emitting most of the \oii.  Since then, the 
{\em HST} images (Hunter \& Thronson 1995, Meurer et al. 
1995, Dufour et al. 1996, 
De Mello et al. 1998) have revealed that the NW region appears 
like a shell of ionized gas about 5\arcsec\ in diameter, encircling a dense 
star cluster. Thus the geometries proposed by Dufour et al. (1988), 
although quite reasonable a priori, do not seem to apply to the case 
under study. 

Campbell (1990), using Ferland's photoionization code CLOUDY, 
constructed 
models to fit the spectral observations of Lequeux et 
al. (1979) obtained through a slit of 3\arcsec.8 $\times$ 12\arcsec.4. These 
observations were giving a \rOiii\ ratio of (3.75 $\pm$ 
0.35) $\times$ 10$^{-2}$. With a 
constant density, ionization bounded spherical model and a LTE Kurucz 
stellar atmosphere with metallicity 1/10 solar, in which the 
adjustable parameters were O/H, \Teff, $n$ and $\epsilon$, Campbell 
obtained 
a best fit model that had \rOiii\ = 3.07~10$^{-2}$, i.e. much lower 
than the 
value she aimed at reproducing. She then proposed a density gradient 
model, in which the inner regions had a density $n$ 
$>$~10$^{5}$\cmcub, so as 
to induce collisional deexcitation of \Oiii. Applying standard 
abundance derivation techniques to this model yields an oxygen 
abundance higher by 70\% than the input value. This led Campbell to 
conclude that I~Zw18 was not as oxygen poor as previously thought. 
The density gradient model of Campbell (1990) can be checked 
directly using the density sensitive \rAriv\ ratio. The only 
observations giving this line ratio are those of Legrand et al. 
(1997), and they indicate a density in the \Arppp\ region lower than 
100\cmcub . Direct images with the {\em HST} do not support Campbell's 
density gradient model either, since, as stated above, 
 the appearance of the \hii\ region 
is that of a shell surrounding the excitation stars.

Stevenson et al. (1993), using a more recent version of CLOUDY, 
constructed a spherical, ionization bounded constant 
density photoionization model to fit their own data. They used as an 
input an extrapolation of the Kurucz LTE model atmospheres. Their 
modelling procedure was very similar to that of Campbell (1990) for 
her constant density model. Their best fit model had O/H = 
1.90~10$^{-5}$ 
and returned \rOiii\ = 2.79~10$^{-2}$, to be compared to their 
observed value of (3.21 $\pm$ 0.42) $\times$ 10$^{-2}$. 

What complicates the discussion of the three studies above is that 
they use different codes with probably different atomic data, and 
they aim at fitting different sets of observations. Nevertheless, it 
is clear that all those
models have difficulties in reproducing the high 
\rOiii\ observed. They have other weak points, as noted by their 
authors. For example, Dufour et al. (1988) and Stevenson et al. (1993) 
comment on the unsatisfactory fitting of the sulfur lines. However, 
the atomic data concerning sulfur are far less well established than 
those concerning oxygen, therefore the discrepancies are not 
necessarily meaningful. Besides, it is not surprising that, with a 
simple density structure, one does not reproduce perfectly at the 
same time the \oiii/\oii\ and \siii/\sii\ ratios. 

The most important defect shared by the three models just discussed 
is that they predict no \Heii\ emission. This is simply due to the 
fact that they used an inadequate input stellar radiation field.
With the presently available stellar population synthesis models for 
the exciting stars of giant \hii\ regions which make use of more 
realistic model atmospheres (Schaerer \& Vacca 1998), and especially 
models that are relevant for the Wolf-Rayet stages of massive stars, 
it is interesting to reanalyze the problem. Using simple photon 
counting arguments, De Mello at al. (1998) have already shown that a 
starburst with a Salpeter initial mass function and an upper mass 
limit of 150\Ms\ could reproduce the equivalent width 
of 
the Wolf-Rayet features and of the narrow \Heii\ emission line in 
I\,Zw\,18. 

It is therefore interesting, using the emergent radiation field from 
such a synthetic stellar population, 
to see whether one can better reproduce the \rOiii\ 
ratio 
observed in I\,Zw\,18, with a model that is more compatible with the 
density structure constrained by the {\em HST} images.

\section{Our model fitting methodology }

\subsection{Computational tools and input parameters}

As in the previous studies, we concentrate on the so-called NW 
component, seen in the top of Fig.\ 1, which shows the WFPC2 
\Ha\ image of the {\em HST} (cf.\ Fig.\ 1 of De Mello et al.\ 1998). 
Throughout the paper, we 
adopt a distance to I~Zw~18 of 10 Mpc, assuming $H_{o}$ = 
75 km~s$^{-1}$~Mpc$^{-1}$, as in many studies (Hunter \& 
Thronson 1995, Martin 1996, van Zee et al. 1998)
\footnote{Izotov et al. (1999) have submitted a paper suggesting 
a distance of 20 Mpc to I Zw 18. Should this be the case, 
the conclusions of our paper that are linked to the ionization structure
and the temperature of the nebula would hardly be changed. The
total mass of the ionized gas would be larger, roughly by a factor 2$^{3}$.}.  

\subsubsection{The stellar population}

We use the same model for the stellar population as described in De 
Mello et al. (1998). It is provided by a evolutionary population 
synthesis code using stellar tracks computed with the Geneva code at 
the appropriate metallicity (1/50\Zs). The stellar atmospheres 
used are spherically expanding non-LTE models for WR stars
(Schmutz et al.\ 1992) and O stars ({\em CoStar} models at $Z=0.004$,
Schaerer \& de Koter 1997), and Kurucz models at [Fe/H]=-1.5 for
the remainder. More details can be found in 
De Mello et al. (1998) and Schaerer \& Vacca (1998).  We assume an 
instantaneous burst of star formation, with an upper mass limit of 
150\Ms\ and a lower mass limit of 0.8 \Ms.
Since all observational quantities considered here depend
only on the properties of massive stars, the choice for 
$M_{\rm low}$ has no influence for the results of this paper.
It merely serves as an absolute normalisation. 
The total initial mass of the stars is adjusted in such a way 
that, at a distance of 10~Mpc, the flux at 3327~\AA\ is equal to 
1.7~10$^{-15}$ \ergsa,
the value measured in the flux calibrated WFPC2 F336W image of De Mello
et al.\ (1998)  within a circle of 2.5\arcsec\ radius centered on the 
NW region (see Fig.\ 1).
This flux is dominated by the latest generation of 
stars in I\,Zw\,18, so that our normalization is hardly sensitive
to the previous star formation history in the NW region of I\,Zw\,18. 
It yields a total stellar 
mass of 8.7~10$^{4}$\Ms, at a distance of 10~Mpc. 

Actually, most of the flux comes from a region much smaller in size, 
and our photoionization modelling is made with the ionizing cluster 
located at the center of the nebula and assuming that its spatial 
extension is negligible. 

We consider that the observed ultraviolet flux is only negligibly 
affected by extinction 
\footnote{A direct fitting of the ultraviolet stellar continuum
by population synthesis models, of which we became aware after the
paper had been submitted, 
yields  C(\Hb) $<$ 0.06 (Mas-Hesse \& Kunth 1999).}. For an extinction C(\Hb) 
of 0.04, such as 
estimated by Izotov \& Thuan (1998), the corrected flux would be only 
about 10\% larger, which is insignificant in our problem. Other 
observers give values of C(\Hb) ranging between 0. and 0.2. If C(\Hb) 
were as large as 0.20, as estimated by  V\'{\i}lchez \& Iglesias-P\'{a}ramo 
(1998) and some other observers, the true stellar flux would be a 
factor two higher. However, all the determinations of C(\Hb), except 
the one by Izotov \& Thuan (1998), do not take into account the 
underlying stellar absorption at \Hb\, and therefore overestimate the 
reddening. 
A further cause of overestimation of C(\Hb), which applies 
also to the work of Izotov \& Thuan (1998), is that the intrinsic 
\Ha/\Hb\ ratio is assumed to be the case B recombination value, while 
collisional excitation of \Ha\ is not negligible in the case of 
I\,Zw\,18 as noted by Davidson \& Kinman (1985). We will come back 
to this below. 

\subsubsection{The nebula}

The photoionization computations are performed with the code PHOTO 
using the atomic data listed in Stasi\'{n}ska \& Leitherer (1996). The 
code assumes spherical geometry, with a central ionizing source. The 
diffuse radiation is treated assuming that all the photons are 
emitted outwards in a solid angle of 2$\pi$, and the transfer of the 
resonant photons of hydrogen and helium is computed with 
the same outward only approximation, but multiplying the photo-absorption 
cross-section by an appropriate factor to account for the increased path 
length due to scattering (Adams 1975). 

The nebular abundances used in the computations are those we derived 
from the spectra of Izotov and Thuan (1998) for the NW component of 
I\,Zw\,18, 
with the same atomic data as used in the photoionization code. For 
helium, however, we adopted the abundance derived by Izotov \& Thuan 
(1998) for the SE component, as stellar absorption contaminates the 
neutral helium lines in the NW component. The nominal value of the 
temperature derived from \rOiii\ is 19800~K. This value was used to 
compute the ionic abundances of all the ions except \Op, \Np\ and 
\Sp, 
for which a value of 15000~K was adopted (this is the typical value 
returned by  our photoionization models for I\,Zw\,18). The electron 
density deduced from \rSii\ is 140\cmcub, and this density was 
adopted in 
the computation of the ionic abundances of all species.  The 
ionization correction factors to compute the total element abundances 
were those of Kingsburgh \& Barlow (1994), which are based on 
photoionization models of planetary nebulae and are also suitable for 
\hii\ regions. They give slightly smaller oxygen abundances (by a few 
\%) than the traditional ionization correction factors which assume 
that the \Oppp\ region is coextensive with the \Hepp\ (we did not 
iterate on the ionization correction factors after our photoionization 
model analysis since this would have not changed any of the conclusions 
drawn in this paper). The carbon 
abundance used 
in the computations follows from the C/O ratio derived by Garnett et 
al. (1997) from {\em HST} observations of I\,Zw\,18. The abundances of the 
elements not constrained by the observations (Mg, Si) and (Cl, Fe) 
have 
been fixed to  10$^{-7}$ and 10$^{-8}$ respectively. Table 1 presents 
the abundance set 
used in all the computations presented in the paper. As already noted 
by previous authors, at the metallicity of I\,Zw\,18, the heavy 
elements 
(i.e. all the elements except hydrogen and helium) play a secondary 
role in the thermal balance. Their role in the absorption of ionizing 
photons is completely negligible. Any change of abundances, even that 
of helium, compatible with the observed intensities of the strong 
lines, will result in a very small change in the electron 
temperature, and we have checked that the effect they will induce in 
the predicted spectra are small compared to the effects discussed 
below.

%\setcounter{table}{0}
%\begin{table*}[tbp]
\begin{table}
\caption{Input abundances relative to hydrogen for the models of 
I\,Zw\,18 (by number).}
\begin{flushleft}
\begin{tabular}{rr}
\hline
He & 7.60 10$^{-2}$\\
C & 3.03 10$^{-6}$ \\
N & 3.89 10$^{-7}$ \\
O & 1.32 10$^{-5}$ \\
Ne & 2.28 10$^{-6}$ \\
S & 3.72 10$^{-7}$ \\
Ar & 9.13 10$^{-8}$ \\
\hline
\end{tabular}
\end{flushleft}
\end{table} 
%\end{table*} 

We do not include dust in the computations. While it is known that, 
in general, dust mixed with the ionized gas may absorb some of the 
ionizing photons, and contribute to the energy balance of the gas by 
photoelectric heating and collisional cooling (e.g. Baldwin et al. 
1991, Borkowski \& Harrington 1991, Shields \& Kennicutt 1995), the expected effect in 
I\,Zw\,18 is negligible, since the dust-to-gas ratio is believed to 
be small at such metallicities (cf.\ Lisenfeld \& Ferrara 1998).

The case of I\,Zw\,18 is thus very interesting for photoionization 
modelling, since due to the very low metallicity of this object, the 
number of unconstrained relevant parameters is minimal.

\subsection{Fitting the observational constraints}

In judging the value of our photoionization models, we do not 
follow the common procedure of producing a table of intensities 
relative to \Hb\ to be compared to the observations. A good 
photoionization model is not only one which reproduces the observed 
line ratios within the uncertainties. It must also satisfy other 
criteria, like being compatible with what is known from the 
distribution of the ionized gas, and what is known of the ionizing 
stars themselves. On the other hand, many line ratios are not at all 
indicative of the quality of a photoionization model. For example, 
obviously, two lines arising from the same atomic level like \Oiii\ 
and [O~{\sc iii}] $\lambda$4959 have intensity ratios that depend 
only on the respective transition probabilities. In \hii\ regions, 
the 
ratio of hydrogen Balmer lines (if case B applies) is little 
dependent on the physical conditions in the ionized gas, and this is 
why it can be used to determine the reddening. The ratios of the 
intensities of neutral helium lines do depend somewhat on the 
electron density distribution and on selective absorption by dust of 
pseudo-resonant photons, (Clegg \& Harrington 1989, Kingdon \& Ferland 
1995), and 
these are introduced in photoionization models. In the 
case of the NW component of I\,Zw\,18, the observed neutral helium lines 
are 
affected by absorption from stars or interstellar sodium 
(Izotov \& Thuan 1998, 
 V\'{\i}lchez \& Iglesias-P\'{a}ramo 1998), and cannot be easily used as 
constraints for 
photoionization models. 

Generally speaking, once line ratios indicative of the 
electron temperature (like \rOiii,  \rNii), of the electron density 
(like 
\rSii, \rAriv) and of the global ionization structure (like 
\Oiii/\Oii\ 
or \Siii/\Sii) have been fitted, the ratios of all the strong 
lines with respect to \Hb\ are necessarily reproduced by a 
photoionization model whose input abundances were obtained from the 
observations. The only condition is that the atomic data to derive 
the abundances and to compute the models should be the same. Problems 
may 
arise only if the empirical ionization factors are different from the 
ones given by the model, or if there is insufficient information on 
the distribution of the electron temperature or density inside the 
nebula (in the case of I\,Zw\,18 no direct information is available
on the temperature in the low ionization zone, but we adopted
 a value inspired by the models).
Therefore, intensity ratios such 
as \Oiii/\Hb, \Neiii/\Hb, \Nii/\Hb, 
\Ariii/\Hb\ or \Ciii/\Hb\ are not a measure of the quality of the 
photoionization model. 

To judge whether a photoionization model is acceptable, one must work 
with outputs that are significantly affected by the physical 
processes on which the photoionization model is based, i.e.
the transfer of the ionizing radiation, the processes determining the 
ionization equilibrium of the various atomic species and the thermal 
balance of the gas. Table 2 lists the quantities that can be used in 
the case of I\,Zw\,18, given the observational information we have on 
the 
object. The value of the \Hb\ flux is derived from the \Ha\ flux
 measured in a circle of radius
$\theta$=2.5\arcsec\ (shown in Fig. 1), 
assuming C(\Hb) = 0. The line 
ratios \Heii/\Hb, \rOiii, \rSii, \Oiii/\Oii, \Siiit/\Sii\ and \Oi/\Hb\ 
are the values observed by Izotov \& Thuan (1998) in the 
rectangular aperture  whose approximate position is shown in Fig. 1. 
It is important to define in 
advance the tolerance we accept for the difference between our model 
predictions and the observations. This must take into account both 
the uncertainty in the observational data, the fact that the spectra 
were taken through an aperture not encompassing the whole nebula, and 
the fact that the nebula does not have a perfect, spherical symmetric 
structure. This latter aspect is, of course, difficult to  quantify, 
and the numbers given in Column 3 of Table 2 are to be regarded 
rather as guidelines. In Column 4, we indicate which is the dominant 
factor determining the adopted tolerance : the signal-to-noise, or 
the geometry. For example, such ratios as \heii/\Hb, \oiii/\oii, \siii/\sii\ or 
\oi/\Hb\ are obviously more dependent on geometrical effects than \rOiii. 
 Note that, even for that ratio,
the tolerance given in Table 2 is larger than the 
uncertainty quoted by Izotov \& Thuan (1998). The reason is that the 
many observations of the NW component of I\,Zw\,18 made over the years, 
with 
different telescopes, detectors, and apertures, yield distinct values 
for this ratio, as shown in Fig. 2. In view of this figure, a 
tolerance of 10\% with respect with the \rOiii\ ratio measured by 
Izotov 
\& Thuan (1998) seems reasonable. The status of the \rSii\ ratio 
is somewhat different. It indicates the average electron density in 
the zone emitting \sii. This is very close to an input parameter, 
since photoionization models are built with a given density 
structure. However, because the density deduced from \rSii\ is not 
the 
total hydrogen density but the electron density 
in the region emitting \sii, and because the 
density is not necessarily uniform, it is important to check that the 
model returns an \rSii\ value that is compatible with the 
observations. 
For the total \Hb\ flux, we accept models giving
 F(\Hb) larger than the observed value, on account 
of the fact that the coverage factor of the ionizing source by the nebula
may be smaller than one.

\setcounter{table}{1}
%\begin{table*}[tbp]

\begin{table*}[htb]
\caption{Observables that a photoionization model of I\,Zw\,18 should 
fit (references in Sect.\ 3.2). }
\begin{flushleft}
\begin{tabular}{llllll}
\hline
%\noalign{\smallskip}                       
%  \multicolumn{4}{c} {(assumed distance 10~Mpc)} \\
Quantity & Value & Tolerance & Major source of uncertainty & Symbol in Figs.\ 4--6 \\
\noalign{\smallskip}                       
\hline
\noalign{\smallskip}                       
F(\Hb) [erg~cm$^{-2}$~s$^{-1}$] & 4.0~10$^{-14}$ & + 0.5 dex & geometry 
(see text) & circle\\
angular radius $\theta$ [arc sec ] & 2.5 & $\pm$ 0.08 dex & geometry & cross \\
\Heii/\Hb  & 0.034 &  $\pm$ 0.2 dex &  
geometry (see text) & square \\
\rOiii\ & 3.28~10$^{-2}$ & $\pm$ 0.04 dex & S/N & open triangle\\
\rSii\ & 1.3 & $\pm$ 0.04 dex & S/N & diamond \\
\Oiii/\Oii\ & 6.82 & $\pm$ 0.1 dex & geometry & filled triangle \\
\Oi/\Hb\ & 0.007  & $\pm$ 0.3 dex & geometry & plus \\
\Siiit/\Sii\ & 0.173 & $\pm$ 0.2 dex & geometry & asterisk \\
\noalign{\smallskip}                       
\hline
\end{tabular}
\end{flushleft}
\end{table*}

Thus, in the following, we compute photoionization models with the 
input parameters as defined in Section 3.1, and see how they 
compare with the constraints specified in Table 2. We will not 
examine the effects of varying the elemental abundances, since, as 
mentioned above, they are negligible in our problem. Uncertainties in 
the measured stellar flux have only a small impact on our models, and 
are therefore not discussed here. 
Similarly, we discard the effects of an error in the distance 
$d$ to I~Zw~18. These are not crucial on the ionization structure of a 
model designed to fit the observed flux at 3327\AA, since 
the mean ionization parameter varies 
roughly like $d^{2/3}$. What we mainly want to see is 
whether, with our present knowledge, we can satisfactorily explain 
the observed properties of I\,Zw\,18. As will be seen, the gas 
density 
distribution plays an important role.

\section{Climbing the ladder of sophistication}

\subsection{The ionizing radiation field }

Before turning to proper photoionization modelling, it is worthwhile 
examining the gross properties of the ionizing radiation field of the 
synthetic stellar population model we are using, and compare it to 
single star model atmospheres. Two quantities are particularly 
relevant. One is  \qhep/\qh, the ratio of the number of photons above 
54.4 and 13.6~eV emitted by the ionizing source. 
This ratio allows one to estimate the \Heii/\Hb\ that would be observed in 
a 
surrounding nebula, by using simple conservation arguments leading to 
the 
formula: \Heii/\Hb\ = 2.14 \qhep/\qh\ (taking the case B 
recombination coefficients given in Osterbrock 1989). As is known, 
this expression is 
valid if the nebula is ionization bounded and the observations 
pertain to the whole volume. It is less commonly realized that it 
also assumes the 
average temperature in the \Hepp\ region to be the same as in the 
entire \hii\ region. This is may be far from true, as will be shown 
below, so a 
correction 
should account for that. Another assumption is that all the photons above 
54.4~eV are absorbed by \Hep\ 
ions. This is not what happens in objects with a low ionization 
parameter. There, the residual neutral hydrogen particles are 
sufficiently numerous to compete with \Hep. In such a case, the 
expression above gives an upper limit to the nebular \Heii/\Hb\ 
ratio. In spite of these difficulties, \qhep/\qh\ remains a useful 
quantity to estimate the intensity of the nebular \Heii\ line. 
Fig. 3a shows the variation of \qhep/\qh\ as a function of 
starburst age for the synthetic model population we are considering. 
As already stated in De Mello et al. (1998), the strength of the 
\heii\ nebular line in I\,Zw\,18 indicates a starburst between 2.9 
and 3.2~Myr. 

Another important ratio is \qhe/\qh, sometimes referred to as the 
hardness parameter of the ionizing radiation field. It provides a 
qualitative 
measure of the heating power of the stars. We have represented this 
quantity in Fig. 3c. We see that, as the 
starburst ages, its heating power gradually declines, and shows only 
a very mild bump at ages around 3~Myr, where the Wolf-Rayet stars are 
present. 
As we will show below this modest increase of the heating power 
is not sufficient to explain the high electron temperature observed 
in I\,Zw\,18. 

For comparison, we show in Figs.\ 3b and d respectively, the values 
of \qhep/\qh\ and \qhe/\qh\ as a function of the stellar effective 
temperature for the LTE model atmospheres of Kurucz (1991), 
the {\em CoStar} model atmospheres corresponding to main sequence 
stars (Schaerer \& de Koter 1997) and for the model atmospheres for 
Wolf-Rayet stars of Schmutz et al.\ (1992). 
The {\em CoStar} models show an increased \Hep\ ionizing flux 
compared to Kurucz models which have a negligible flux even for 
very low metallicities ([Fe/H]$=-1.5$). The reasons for this
difference have been discussed in Schaerer \& de Koter (1997).
In addition to the \teff\ dependence,  \qhep\ from WR models depend 
strongly on the wind density. \Hep\ ionizing photons are only 
predicted by models with sufficiently thin winds (cf.\ Schmutz et al.\ 
1992).
Figure 3d shows the increase of the hardness of the radiation field,
at a given \teff, 
between the spherically expanding non-LTE models for O and WR stars
and the traditional Kurucz models (see discussion in Schaerer \& de 
Koter 1997). This provides a greater heating power which, as will
be shown later, is however still insufficient to explain the 
observations.

\subsection{I\,Zw\,18 as a uniform sphere}

We start with the easiest and most commonly used geometry in
photoionization modelling: a  sphere uniformly filled with 
gas at constant density, occupying a fraction $\epsilon$ of the whole 
nebular volume. The free parameters of the models are then only the 
age of the starburst, the gas density and the filling factor. Each 
model is computed starting from the center, and the computations are 
stopped either when the \oiii/\oii\ ratio has reached the observed 
value given in Table 2, or when the gas becomes neutral. In other 
words, we examine also models that are not ionization bounded, in 
contrast to previous studies.

Figure 4 shows our diagnostic diagram for a series of models having a 
density $n$= 100\cmcub\ and a filling factor $\epsilon$=0.01. The left 
panel shows the 
computed values of log F(\Hb) + 15 (open circles), log \Heii/\Hb\ + 2 
(squares), angular radius $\theta$ (crosses), 
\rOiii\ $\times$ 100 (open triangles), \rSii\ (diamonds), 
log (\oiii/\oii)
(black triangles), log (\siii/\sii) (asterisks) and log \oi/\Hb\ +3
(plus)  as a 
function of the starburst age. The black circles correspond to the 
value of log F(\Hb) + 15  that the nebula would have if it were 
ionization bounded. Thus, by comparing the positions of an open 
circle and a black circle, at a given abscissa, one can immediately 
see whether
the model is density bounded and how much diffuse \Ha\ or \Hb\ emission
 is expected to be emitted outside the main body of the nebula.
 In the right panel, the observed 
values are represented on the same vertical scale and with the same 
symbols as the model predictions. The tolerances listed in Table 
2 are represented as vertical error bars (the horizontal displacement 
of the symbols has no particular meaning). We readily see that the 
age of the starburst is 
important only for the \Heii\ line, the other quantities varying very 
little for ages 2.7--3.4 Myr. Therefore, for the following 
runs of models, we adopt an age of 3.1~Myr. In principle, one 
can always adjust the age for the model to reproduce the 
observed \Heii/\Hb\ ratio exactly. 

Figure 5 shows the same sort of diagnostic diagram as Fig. 4 for a 
series of models with $n$ = 100\cmcub\ and varying filling factor. 
For a 
filling factor around 0.1 or larger, with the adopted electron density, 
the model is ionization bounded, and its \oiii/\oii\ is larger than 
observed. For filling factors smaller than that, the gas distribution 
is more extended, so that the general ionization level drops. The 
observed \oiii/\oii\ can then only be reproduced for a density 
bounded 
model. In such a case, the \Hb\ radiation actually produced by the 
nebula is smaller 
than if all the ionizing photons were absorbed in the nebula. 
A filling factor of 0.002 -- 0.05 gives values of \oiii/\oii, F(\Hb) 
and $\theta$  in agreement with the observations.
 But such models give \rOiii\ too small 
compared with the observations, and \oi/\Hb\ below the observed 
value by nearly two orders of magnitude. It is interesting, though,  
to understand the qualitative behavior of these line ratios as 
$\epsilon$ decreases. \oi/\Hb\ decreases because the model becomes 
more 
and more density bounded in order to reproduce the observed 
\oiii/\oii\, and levels off at $\epsilon$ = 0.1, because the 
ionization 
parameter of the model is then so small that the \oi\ is gradually 
emitted by residual neutral oxygen in the main body of the 
nebula and not in the outskirts. \rOiii\ decreases as $\epsilon$
decreases, because of the increasing proportion of L$\alpha$ cooling 
as the ionization parameter drops. 

One can build other series of models with different values of $n$ 
that are still compatible with the observed \rSii. Qualitatively, 
their behavior is 
the same and no acceptable solution is found.

Interestingly, Fig. 5 shows that models  with $\epsilon$  $\ge$ 0.1 
have \rOiii\ marginally compatible with the observations ( \rOiii\ = 3.03 
10$^{-2}$ for $\epsilon$ = 0.1), but such models have too large 
\oiii/\oii ($>$ 10 compared of the observed value 6.8) and too small angular 
radius ($<$ 1.6\arcsec\ instead of the observed value 2.5\arcsec). Note, 
by the way, that such models, being optically thick, return a rather 
large \oi/\Hb, actually close to the observed value, and a 
\siii/\sii\ compatible with the observations. However, we do not 
use \siii/\sii\ as a primary criterion to judge the validity of a model, 
since experience with photoionization modelling of planetary nebulae shows 
that it is difficult to reproduce at the same time the sulfur and the 
oxygen ionization structure of a given object, and, in principle, one 
expects the atomic data for oxygen to be more reliable than those for 
sulfur. The strongest argument against models with $\epsilon > 0.1$ is 
their angular size, which is definitely too small compared with the 
observations. This remains true even when considering a reasonable error 
on the distance since, with the condition that we impose on the flux at 
3327~\AA\ to be preserved,  the angular radius of a model goes roughly 
like $d^{1/3}$. This illustrates the importance of taking into account 
other parameters in addition to line ratios to accept or reject a 
photoionization model.
 
\subsection{I~ Zw~18 as a spherical shell }

The series of models presented above had mainly a pedagogical 
value, but they are obviously 
incompatible with the observed morphology in \Ha. The next step is 
to consider a model consisting of a hollow, spherical shell of 
constant density, similar to the one constructed by  Garc\'{\i}a-Vargas et al. (1997) 
for NGC\,7714 for example. In such a case, there is an additional free parameter,
$R_{in}$, the radius of the inner boundary of the shell. It is fixed, 
more or less, by the appearance of the \Ha\ image. Figure 6 shows a 
diagnostic diagram for a series of models with $R_{in}$ = 2.25 
10$^{20}$~cm (corresponding to an angular radius of 1.5\arcsec), $n$ = 
100\cmcub\, and varying $\epsilon$. The qualitative behavior is 
similar to 
that seen for the uniform sphere models presented in Fig. 1, but the 
\rOiii\ ratio is now even lower (it never exceeds 2.5~10$^{-2}$ in 
this series). This is because of the enhanced role 
of 
L$\alpha$ cooling, which is strong in all the parts of the nebula, 
while for the full sphere model, in the zone close to the star, the 
ionization parameter is very high and consequently the population of 
neutral hydrogen very small. 

Apart from the \rOiii\ problem, models with $\epsilon$ = 0.002 -- 
0.05 are satisfactory as concerns the main diagnostics (\oiii/\oii,
F(\Hb) and $\theta$). The models become progressively 
density bounded 
towards smaller values of $\epsilon$ $<$ 0.02, meaning that there is 
a leakage of ionizing photons. From Fig. 6, one sees that 
these photons are enough to produce an \Ha\ emission in an extended diffuse 
envelope that is at least comparable in strength to the total 
emission from the dense shell. This is in agreement with Dufour \& 
Hester's (1990) ground-based observation of extended \Ha\ emission 
surrounding the main body of star formation. 

\subsection{Other geometries}
Closer inspection of the {\em HST} \Ha\ image shows that the gas emission
is incompatible with a spherical bubble.
This is illustrated in Fig.\ 7, where the observed cumulative
surface brightness profile within radius $r$ (dashed line) is compared 
to the expected profiles for constant density spherical shells of
various inner radii (solid lines).
The theoretical profiles are  
obtained assuming that the temperature is uniform 
in the gas, but taking into account a reasonable temperature gradient 
in the model hardly changes the picture. Clearly, the observed 
profile is not compatible with a spherical density distribution. The 
column density of emitting matter in the central zone of the image is 
too small. One must either have an incomplete shell 
with some matter stripped off from the poles, or even a more extreme 
morphology like an diffuse axisymmetric 
body with a dense ringlike equator seen face on. Such geometries are 
actually common among planetary nebulae (Corradi \& Schwartz 1995) 
and nebulae surrounding luminous blue variables (Nota et al. 1995), 
being probably the result of the interaction of an aspherical  
stellar wind from the central stars (Mellema 1995, Frank et al. 1998) 
and are also suggested to exist in superbubbles and supergiant shells 
and to give rise to the blow-out phenomenon (Mac Low et al. 1989, 
Wang \& Helfand 1991, Tenorio-Tagle et al. 1997, Oey \& Smedley 1998,
 Martin 1998).

Does the consideration of such a geometry help in solving the \rOiii\ 
problem? In the case of a spherical shell with some matter lacking at 
the poles, our 
computations overestimate the role of the diffuse ionizing radiation, 
which is supposed to come from a complete shell. Since the 
heating power of the diffuse ionizing radiation is smaller than that 
of the stellar radiation, one may be underestimating the electron 
temperature. As a test, we have run a model switching off completely 
the diffuse radiation, so as to maximize the electron temperature. 
The effect was to increase the \Opp\ temperature by only 200~K. In 
the case of a ring with some diffuse matter seen in projection inside 
the ring, the gas lying close to the stars would be 
at a higher electron temperature than the matter of the ring, and one 
expects that an adequate combination of the parameters describing the 
gas distribution interior to and inside of the ring
might reproduce the \rOiii\ measured in the 
aperture shown in Fig. 1. However, we notice that the region of high 
\rOiii\ is much larger than that. It extends over almost 20\arcsec\ (
 V\'{\i}lchez \& Iglesias-P\'{a}ramo 1998) and \rOiii\ is  
neither correlated with \oiii/\oii\ nor with the \Ha\ surface 
brightness. Therefore, one cannot explain the high \Oiiit\ by the 
emitting matter being close to the ionizing stars.
In passing we note that the observations of  V\'{\i}lchez \& 
Iglesias-P\'{a}ramo (1998) show the nebular \Heii\ emission to be 
extended as well (although not as much as \Oiiit). If this emission is 
due to photoionization by the central star cluster, as modelled in 
this paper, this means that the \Ha\ ring is porous, since \Heii\ 
emission necessarily comes from a region separated from the stars by 
only a small amount of intervening matter.

In summary, we must conclude that, if we take into account all the 
available information on the structure of I\,Zw\,18, we are not able 
to 
explain the high  observed \rOiii\ ratio with our photoionization 
models. In our best models, \rOiii\ is below the nominal value of 
Izotov \& Thuan (1998) by 25 - 35\%.

\subsection{Back to the model assumptions}

Can our lack of success be attributed to an improper description of 
the stellar radiation field? After all, we know little about the 
validity of stellar model atmospheres in the Lyman continuum
(see discussion in Schaerer 1998).
Direct measurements of the EUV flux of early B stars revealed an 
important EUV excess (up to $\sim$ 1.5 dex) with respect to plane 
parallel model atmospheres (Cassinelli et al.\ 1995, 1996), whose 
origin has been discussed by Najarro et al.\ (1996), Schaerer \& de 
Koter (1997) and Aufdenberg et al.\ (1998).
For O stars a similar excess in the {\em total} Lyman continuum
output is, however, excluded from considerations of their bolometric 
luminosity and measurements of \hii\ region luminosities 
(Oey \& Kennicutt 1997, Schaerer 1998).
The hardness of the radiation field, which is crucial for the heating
of the nebula, is more difficult to test.
Some constraints can be obtained by comparing the line emission of 
nebulae surrounding hot stars with the results of photoionization models
(Esteban et al.\ 1993,  Pe\~{n}a et al.\ 1998, Crowther et al.\ 1999),
but this is a difficult task, considering that 
the nebular emission depends also on its geometry. 
Although the hardness predicted by  the {\em CoStar} O stars models 
permits to build grids of photoionization models that seem to explain the 
 observations of Galactic and LMC \hii\ regions 
(Stasi\'nska \& Schaerer 1997), the constraints are not sufficient
to prove or disprove the models.

To check the effect of a harder radiation field, we have run a series of models 
where the radiation field above 24.6~eV was arbitrarily multiplied by 
a factor 3 (raising the value of \qhe/\qh\ from  0.33 to 0.59,
corresponding to \teff\ from $\ga$ 40000 K to $\sim$ 100\,000~K or
a blackbody of the same temperature (cf.\ Fig.\ 3d).
This drastic hardening of the radiation field resulted in an increase 
of \rOiii\ of only 10\%. It is only by assuming a blackbody radiation 
of 300\,000~K (which has \qhe/\qh\ = 0.9) that one approaches the 
observed \rOiii. A model similar to those presented in Fig. 6 but 
with such a radiation field gives \rOiii\ =3.15 10$^{-2}$. But is has 
a \Heii/\Hb\ ratio of 0.53, which is completely ruled out by the 
observations. Of course, a blackbody is probably not the best 
representation for the spectral energy distribution of the radiation 
emitted by a very hot body, but in order to explain the emission line 
spectrum of I~Zw~18 by stars, one has to assume an ionizing radiation 
strongly enhanced at energies between 20 -- 54 eV, but not above 54.4 
eV, compared to the model of De Mello et al.\ (1998). 
If this is realistic cannot be said at the present time.

We have also checked the effect of heating by additional X-rays that 
would be emitted due to the interaction of the stellar winds with 
ambient matter (see Martin 1996 for such X-ray observations),
 by simply adding a bremsstrahlung spectrum at T = 10$^{6}$ 
or  T = 10$^{7}$ K to the radiation from the starburst model. As 
expected, the effect on \rOiii\ was negligible, since the X-rays are 
mostly absorbed in the low ionization regions (they do raise the 
temperature in the \Op\ zone to T$_{e}$ $\simeq$ 16\,000~K). 

As already commented by previous authors, changing the elemental 
abundances does not improve the situation. Actually, even by putting 
all the elements heavier than helium to nearly zero abundance, 
\rOiii\ is 
raised by only 7\%. Varying the helium abundance in reasonable limits 
does not change the problem.

The neglect of dust is not expected to be responsible for the
too low \rOiii\ we find in our models for I~Zw~18.
Gas heating by photoemission from 
grains can contribute by as much as 20\% to the electron thermal 
balance when the dust-to-gas ratio is similar to that found in the 
Orion nebula. But, as discussed by Baldwin et al. (1991), it is 
effective close to the ionizing source where dust provides most of 
the opacity. Besides, the proportion of dust in I\,Zw\,18 is expected
to be small, given the low metallicity. Extrapolating the relation found by 
Lisenfeld \& Ferrara (1998) between the dust-to-gas ratio and the 
oxygen abundances in dwarf galaxies to the metallicity of I\,Zw\,18 
yields a dust-to-gass mass ratio 2 to 3 orders of magnitudes smaller 
than in the solar vicinity.

It remains to examine the ingredients of the photoionization code. We 
first stress that the \rOiii\ problem in I\,Zw\,18 has been 
encountered 
by various authors using different codes, even if the observational 
material and the points of emphasis in the discussion were not the 
same in all the studies.
As a further test, we compared the same model for I~Zw~18 run by CLOUDY 90
and by PHOTO and the difference in the predicted value of \rOiii\ was 
only 5\%.
One must therefore incriminate 
something that is wrongly treated in the same way in many codes. 
One possibility that comes to mind is the diffuse ionizing radiation,
which is treated by  some kind of outward only approximation in all
codes used to model I~Zw~18.
However,  we do not expect 
 that an accurate treatment of the diffuse ionizing radiation 
would solve the problem.
Indeed, comparison of codes that treat more 
accurately the ionizing radiation with those using an outward only 
approximation shows only a negligible difference in the electron 
temperature (Ferland et al.\ 1996). 
Besides, as we have shown, even quenching the diffuse ionizing 
radiation does not solve the problem. 

Finally, one can also question the 
atomic data. The most relevant ones here are those governing the 
emissivities of the observed \oiii\ lines and the \hi\  collision 
strengths. The magnitude of the discrepancy we wish to solve would 
require modifications of the collision strengths or transition 
probabilities for \oiii\ of about 25\%. 
This is much larger than the expected uncertainties and the differences 
between 
the results of different computations for this ion (see discussion in 
Lennon and Burke 1994 and  Galavis et al. 1997). Concerning L$\alpha$ 
excitation, dividing the collision strength by a factor 2 (which is 
far above any  conceivable uncertainty, see e.g. Aggarwal et al. 
1991) modifies \rOiii\ only by 2\% because L$\alpha$ acts like a 
thermostat.

We are therefore left with the conclusion that the \rOiii\ ratio 
cannot be explained in the framework of photoionization models alone. 

\subsection{Condensations and filaments}

Another failure of our photoionization models is that 
they predict too low \oi/\Hb\ compared to observations. This is a 
general feature of photoionization models and it is often taken as 
one of the reasons to invoke the presence of shocks. However, it is 
well known that the presence of small condensations of filaments 
enhances the \oi/\Hb\ ratio, by reducing the local ionization 
parameter. Another possibility is to have an intervening filament 
located at a large distance from the ionizing source, whose projected 
surface on the aperture of the observations would be small. 

In order to see under what conditions such models can quantitatively 
account for 
the observed \oi/\Hb\ ratio, we have mimicked such a situation by 
computing a series of ionization bounded photoionization 
models for filaments of different densities located at various distances 
from the exciting stars. 
For simplicity, we assumed that there is no intervening matter 
between the filaments and the ionizing source.
The models were actually computed 
for complete shells. The radiation coming from a filament 
can be simply obtained 
by multiplying the flux computed in the model by the 
covering factor $f$ by 
which the filament is covering the source. 

\begin{table*}
\caption{Line intensities relative to \Hb\ for models of filaments.
}
\begin{flushleft}
\begin{tabular}{l|l|lll|ll}
\hline
\noalign{\smallskip}  
 &     Main body & \multicolumn{3}{l|}{Filaments of various $n$} &  
 \multicolumn{2}{l}{Filaments at various distances} \\
\hline
\noalign{\smallskip} 
$n$ [cm$^{-3}$]  & 10$^2$   $^a$   &   10$^4$    &  10$^5$  &  10$^6$ & 10$^2$ &  10$^2$   \\
$\theta_{\rm in}$ [\arcsec] &  &       1.5   &   1.5    &   1.5   &   20.  &  100.    \\
\noalign{\smallskip}                       
\hline
\noalign{\smallskip}                       
F(\Hb)$^b$
 & 2.02e-13   &   3.33e-13  & 3.32e-13 & 3.35e-13  & 3.41e-13 &  
3.47e-13 \\
\Oi\   & 1.04e-4    &    5.28e-2  & 1.57e-1  & 3.83e-1   & 6.90e-2  &  
2.81e-1  \\
%3727.+
\Oii\  & 2.15e-1    &    2.41e-1  & 5.39e-2  & 1.19e-2   & 4.71e-1  &  
2.85e-1  \\
%4363.
\Oiiit\   & 3.46e-2    &    3.63e-5  & 1.12e-6  & 1.04e-7   & 1.20e-5  &  
1.74e-7  \\
%5007
\Oiii\    & 1.42       &    3.76e-3  & 1.41e-4  & 1.25e-5   & 1.41e-3  &  
3.08e-5  \\
\sii\ $\lambda$6717 & 5.58e-4    &    5.20e-2  & 5.23e-2  & 5.27e-2   & 1.87e-1  &  
5.70e-1  \\
\sii\ $\lambda$6731.   & 4.19e-3    &    8.04e-2  & 9.16e-2  & 1.03e-1   & 1.35e-1  &  
3.96e-1  \\
\noalign{\smallskip}                       
\hline
\noalign{\smallskip}                       
\noalign{\small
$^a$ The model for the main body is density bounded (see text)}
\noalign{\small
$^b$ The total \Hb\ flux is given in  erg s$^{-1}$ for a covering factor $f$=1. }
\end{tabular}
\end{flushleft}
\end{table*}

Table 3 presents the ratios with respect to \Hb\ of the \Oi, \Oii, 
\Oiii, [S~{\sc ii}] $\lambda$6717 and [S~{\sc ii}] $\lambda$6731 for 
these models. The first column of the table corresponds to a 
photoionization model for the main body of the nebula with  
$R_{in}$ = 2.25 10$^{20}$ cm (corresponding to  
$\theta_{\rm in}=$ 1.5\arcsec), $n$=100\cmcub, and $\epsilon$=0.01, 
and  density bounded so 
as to obtain the observed \oiii/\oii\ (this is one of the models of 
Fig. 6). It can be readily seen that, for an intervening filament of 
density 10$^{2}$\cmcub\ located at a distance of 500~pc from the star 
cluster, or for condensations of density 10$^{6}$\cmcub, one can 
reproduce the observed \oi/\Hb\ line ratio without 
strongly affecting the remaining lines, not even the density 
sensitive \rSii\ ratio, if one assumes a covering factor 
$f$ of about 0.1.

This explanation may appear somewhat speculative.
 However, one must 
be aware that the morphology of the ionized gas in I\,Zw\,18 shows 
filaments in the main shell as well as further out and this has been 
amply discussed in the literature (Hunter \& Thronson 1995, Dufour
et al. 1995, Martin 1996). Our point is that, even in the 
framework of pure photoionization models, if one accounts for a 
density structure suggested directly by the observations, the 
strength of the \oi\ line can be easily understood. 

We note that the condensations or filaments that produce \oi\  are 
optically thick, and therefore their 
neutral counterpart should be seen in \hi. Unfortunately, the 
available \hi\ maps of I\,Zw\,18 (van Zee et al. 1998) do not have 
sufficient spatial resolution to reveal such filaments. But these 
authors show that the peak emission in \hi\ and \Ha\ coincide (in 
their paper, 
the peak emission in \Ha\ actually refers to the whole NW component), and 
find that the entire optical system, including the diffuse emission, 
is embedded in a diffuse, irregular and clumpy neutral cloud. In such 
a situation, it is very likely that some clumps or filaments, 
situated in front of the main nebula, and having a small covering 
factor, produce the observed \oi\ line. 

By using photoionization 
models to explain the observed \oi/\Hb\ ratio, one can deduce 
the presence of density fluctuations, even if those are not seen 
directly.  Such density fluctuations could then be, more directly, 
inferred from density sensitive ratios of [Fe~{\sc ii}]  lines, such 
as seen 
and analyzed in the Orion nebula by Bautista \& Pradhan (1995).

\subsection{A few properties of I\,Zw\,18 deduced from models}

Although we have not built a completely satisfactory photoionization 
model reproducing all the relevant observed properties of I\,Zw\,18, 
we 
are not too far from it. We describe below some properties of the 
best models that may be of interest for further empirical studies of 
this object. 

Tables 4 and 5 present the mean ionic fractions and mean ionic 
temperatures 
(as defined in Stasi\'{n}ska 1990) for the different elements 
considered, in the case of the best fit models with a uniform sphere 
and a spherical shell respectively, with $n$=100\cmcub, and $\epsilon$=0.01.

It can be seen that, while both geometries yield similar ionic 
fractions for the main ionization stages, the relative populations of 
the highly charged 
trace ions are very different. In the case of the uniform sphere, 
the proportion of \Oppp, for example, is twice as large as in 
the shell model. Also, the characteristic temperature of ions with 
high ionization potential are much higher in the case of the filled 
sphere, for reasons commented on earlier. As a result, the 
OIV]1394/\Hb\ ratio is 9.7~10$^{-3}$ in the first case and 8.6~10$^{-4}$ 
in the second. This line is too weak to be measured in I\,Zw\,18, of 
course, but it is useful to keep this example in mind for the study 
or more metal rich objects. 

It is interesting to point out that, in the case of the uniform 
sphere, the total flux in the \Heii\ line is smaller than in the case 
of the shell model (3.1~10$^{-15}$  vs. 3.3~10$^{-15}$ erg~cm$^{-2}$~s$^{-1}$)
 despite the 
fact that the ionic fractions of \Hepp\ are similar. This is 
because the \Hepp\  region is at a much higher temperature (25\,000~K 
versus 18\,000~K).

Tables 4 and 5 can be used for estimating the ionization correction 
factors for I\,Zw\,18.
Caution should, however, be applied regarding especially
the ionization structure predicted for elements of the third row of 
Mendeleev's table.  
Experience in photoionization modelling of 
planetary nebulae, where the observational constraints are larger, 
shows that the ionization structure of these elements is rarely 
satisfactorily reproduced with simple models (Howard et al. 1997, Pe\~na 
et al., 1998).

The total ionized mass in the NW component is relatively well determined, 
since we know 
the total number of ionizing photons, the radius of the emitting 
region and have 
an estimate of the 
mean ionization parameter through the observed \oiii/\oii. Indeed,
at a given \qh,  
for a constant density sphere with a filling factor $\epsilon$, 
$n^{2} \epsilon$ is proportional to the cube of the radius,
 while $n \epsilon^{2}$ is proportional to the cube of the ionization 
 parameter. Of course, we have just made the point that the NW 
 component
 of I\,Zw\,18 is not a sphere. Nevertheless, we have an order of 
 magnitude estimate, which is of 3.~10$^{5}$\Ms\ at $d$ = 10~Mpc (this 
 estimate varies like $d^{3}$). 

Finally, it is important to stress, as already mentioned above, that 
\Ha\ is partly excited by collisions. In all our models for the main 
body of the nebula,  \Ha/\Hb\ lies between 2.95 and 3, 
while the case B recombination value is 2.7. This means that the 
reddening of I\,Zw\,18 is smaller than the value obtained using the 
case B 
recombination value at the temperature derived from \rOiii. If we 
take the observations of Izotov \& Thuan (1998), who also correct for 
underlying stellar absorption, we obtain C(\Hb)=0.

\section{Summary and conclusions}

We have built photoionization models for the NW component of I~Zw~18 
using the radiation field from a starburst population 
synthesis at appropriate 
metallicity (De Mello et al. 1998) that is consistent with the 
Wolf-Rayet signatures seen in the spectra of I~Zw~18. The aim was to 
see whether, with a nebular density structure compatible with recent 
{\em HST} images, it was possible to explain the high 
 \rOiii\ ratio seen in this object, commonly interpreted as 
 indicative of electron temperature $\simeq$ 20\,000K.

For our photoionization analysis we have focused on properties which
are relevant and crucial model predictions. 
For the observational constraints we have not only used line ratios, but 
also other properties such as the integrated stellar flux at 3327~\AA\ 
and the observed angular radius of the emitting region as seen by the
{\em HST}. Care has also been taken to include tolerances on model 
properties which may be affected by deviations from various simple 
geometries.

We have found that \rOiii\ cannot be explained by pure photoionization 
models, which yield too low an electron temperature. We have 
considered the effects due to departure from spherical symmetry 
indicated by the {\em HST} images. 
Indeed these show that 
the NW component of I~Zw~18 is neither a uniform 
sphere nor a spherical shell, but rather a bipolar structure with a 
dense equatorial ring seen pole on. 
 
We have discussed the consequences that an inaccurate description of the 
stellar ionizing radiation 
field might have on \rOiii\, as well as additional photoionization 
by X-rays.
Finally, we have considered possible errors in the atomic data.

All these trials were far from solving the electron temperature 
problem, raising the \rOiii\ ratio by only a few percent while the 
discrepancy with the observations is on the 30\% level. 
Such a discrepancy means that we are missing a heating source whose 
power may be of the same magnitude as that  of the stellar ionizing 
photons. It is also possible that 
the unknown energy source is not so powerful, but acts in such a way 
that small quantities of gas are emitting at very high temperatures, 
thus boosting the \Oiiit\ line. Shocks are of course one of the 
options (Peimbert et al. 1991, Martin 1996, 1997), as well as conductive heating 
at the interface of an X-ray plasma with optically visible gas (Maciejewski et al. 1996). 
Such ideas need to be examined quantitatively, and applied to the 
case of I~Zw~18, which we shall attempt in a future work.

What are the consequences of our failure in understanding the energy 
budget on the abundance determinations in I~Zw~18? It depends on how 
 the electron temperature is distributed in the \Opp zone. 
 As emphasized by Peimbert (1967, 1996) over the years (see also 
 Mathis et al. 1998), the existence of even small zones at very high
  temperatures will boost  the lines with high excitation threshold 
  like \Oiiit, so that the temperature derived from \rOiii\ will 
  overestimate the average temperature of the regions emitting \Oiii. 
  Consequently, the true O/H ratio will be larger than the one derived 
  by the standard methods. The C/O ratio, on the other hand,  will be 
  smaller than derived empirically, because the ultraviolet \Ciii\ 
  line will be extremely enhanced in the high temperature regions. 
  Such a possibility was invoked by Garnett et al. (1997) to explain 
  the high C/O found in I~Zw~18 compared to other metal-poor 
  irregular galaxies. Presently, however, too little is known both 
  theoretically and observationally to estimate quantitatively this
effect, 
  and it is not excluded that the abundances derived so far may be 
  correct within 30\%. 
  Obviously, high spatial resolution mapping of the \rOiii\ ratio in 
  I~Zw~18 would be valuable to track the origin of the high \Oiiit\ 
  seen in spectra integrated over a surface of about 10~\arcsec $^{2}$.
  
Beside demonstrating the existence of a heating problem in I~Zw~18, 
our photoionization model analysis led to several other results.

The intensity of the nebular \Heii\ line can be reproduced with a 
detailed photoionization 
model having as an input the stellar radiation field that is consistent
the observed Wolf-Rayet features in I~Zw~18.  
This confirms the results of De Mello et al.  (1998) based on simple
Case B recombination theory.

By fitting the observed \oiii/\oii\ ratio and the angular size of the NW 
component with a model where the stellar radiation flux was adjusted 
to the observed value, we were able to show that the \hii\ region is 
not ionization bounded and about half of the 
ionizing photons are leaking out of it, sufficient to explain the extended 
diffuse \Ha\ emission observed in I~Zw~18. 

While the \oi\ emission is not reproduced in simple models, it can 
easily be accounted for by condensations or by intervening filaments
on the line of sight. There is no need to invoke shocks to excite the 
\oi\ line, although shocks are probably involved in the creation of 
the filaments, as suggested by Dopita (1997) in the context of 
planetary nebulae.

The intrinsic \Ha/\Hb\ ratio is significantly affected by collisional
 excitation: our photoionization models give a value of 3.0, to be compared
  to the case
 B  recombination value of 2.75 used in most observational papers. 
 Consequently, the reddening is smaller than usually estimated, with 
 C(\Hb) practically equal to 0. 

Our models can be used to give ionization correction factors 
appropriate for I~Zw~18 for more accurate abundance determinations. 
However, the largest uncertainty in the abundances of C, N, O and Ne 
ultimately lies in the unsolved temperature problem.

It would be, of course, of great interest to find out whether other 
galaxies share with I~Zw~18 this \rOiii\ problem. 
There are at least two other cases which would deserve a more 
thorough analysis.
One is the starburst galaxy NGC 7714, for which
published photoionization models (Garc\'{\i}a-Vargas et al. 1997) 
also give \rOiii\
smaller than observed. However, it still needs to be demonstrated 
that this problem remains when modifying the model assumptions
 (e.g. the gas density distribution, possible heating of the
gas by photolelectric effect on dust particles etc...).
In the case of NGC 2363, the photoionization models of
Luridiana et al. (1999) that were built using the oxygen abundances
derived directly from the observations yielded a \rOiii\ ratio marginally
compatible with the observations. These authors further argued that,  
due to the presence of large spatial temperature fluctuations, 
the true gas metallicity  in this object is higher than derived by 
empirical methods. In such a case, the \rOiii\ ratio becomes even more
discrepant.

It might well be that additional
heating sources exist in giant \hii\ regions, giving rise to such
large temperature variations and and enhancing the \Oiiit\ emission. 
As mentioned above, such
a scenario needs to be worked out quantitatively.

Further detailed observational and theoretical studies of individual objects 
would be helpful,
since we have shown that with insufficient observational constraints, 
high \rOiii\ may actually be produced by photoionization models. 
The effort is worthwhile, since it 
would have implications both on our understanding of the energetics of 
starburst galaxies and on our confidence in abundance derivations.

%\appendix{}

\begin{acknowledgements}
This project was partly supported by the ``GdR Galaxies''.
DS acknowledges a grant from the Swiss National Foundation of Scientific
Research.
We thank Duilia De Mello for providing the {\em HST} images and
Jean-Fran\c{c}ois Le Borgne for help with IRAF.
During the course of this work, we benefited from conversations 
with Jose V\'{\i}lchez, Rosa  Gonz\'{a}lez-Delgado, Enrique P\'{e}rez, 
Yurij Izotov, Trinh Xuan Thuan. Thanks are due to Valentina Luridiana, 
Crystal Martin and Claus Leitherer for reading the manuscript.
\end{acknowledgements}

\begin{figure}[htb]        
\centerline{\psfig{figure=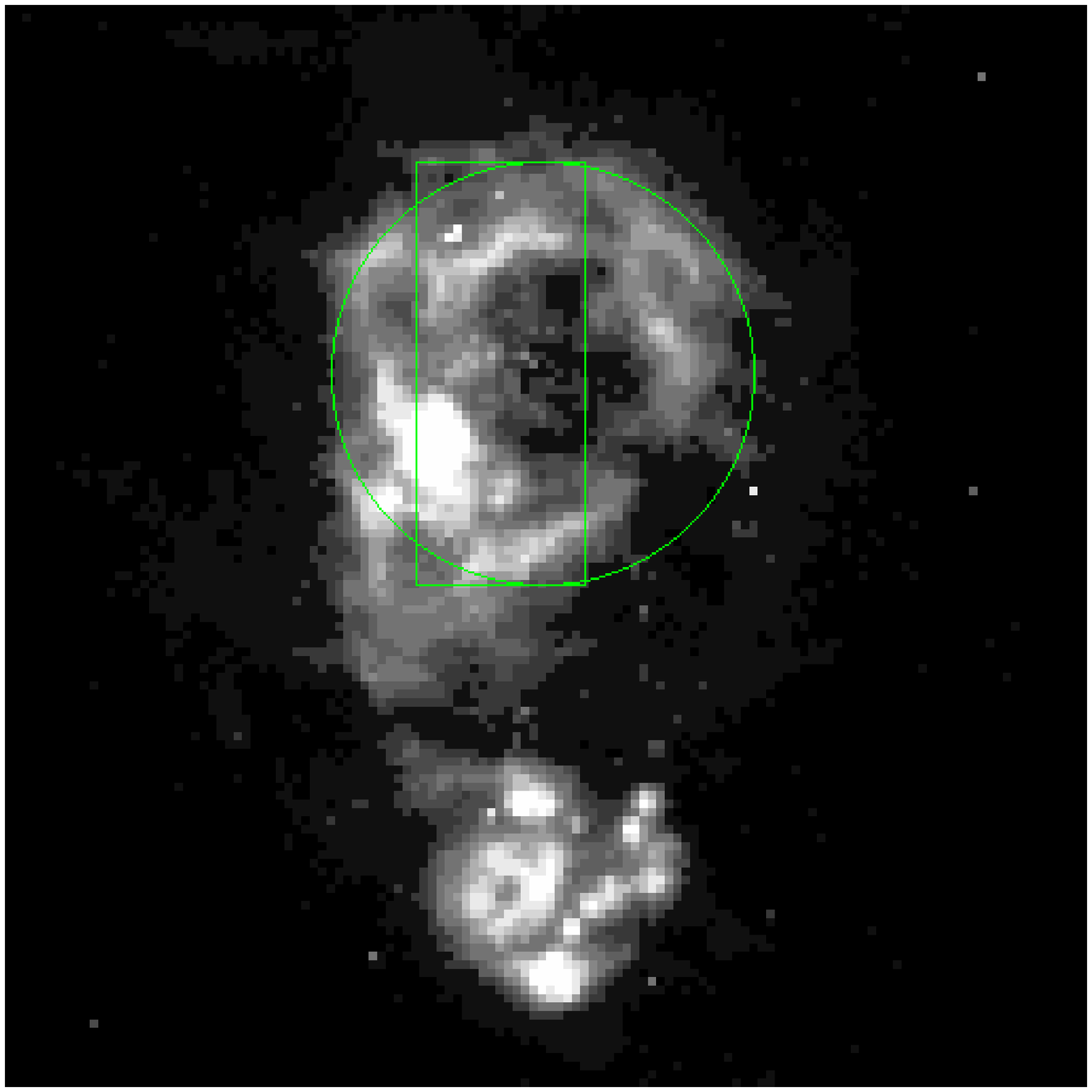,width=8.8cm}}
\caption{WFPC2 \Ha\ {\em HST} image of I~Zw~18 aligned on NW-SE knots
(WF3 image transformed to PC resolution, cf.\ Fig.\ 1 from De Mello 
et al.\ 1998; upper region: NW, lower region: SE)
The circle (radius 2.5\arcsec) denotes the area used for the measurements
of the \Ha\ and F336W flux (cf.\ Sect.\ 3.1.1, Table 2).
The rectangle illustrates the aperture used for the 
spectroscopic observations of Izotov \& Thuan (1998) centered on the NW knot.
The absolute position is indicative only.
F336W and F555W images indicate a small shift ($\sim$ 0.5--1 \arcsec) 
between the peak emission and the center of the \Ha\ emission
(as shown by the circle). A shift of 0.5 \arcsec\ was used
for the aperture of Izotov \& Thuan (1998).
}
\label{HST images (V,Halpha) with slit positions}
\end{figure}

\begin{figure}[htb]        
\centerline{\psfig{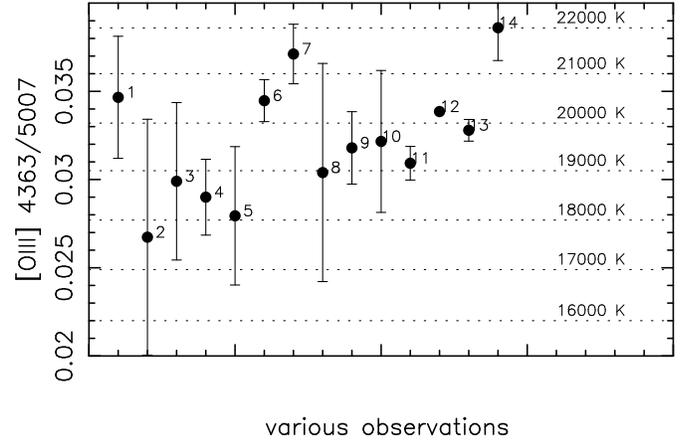}}
\caption{Observed \rOiii\ ratios in I~Zw~18 and corresponding 
electron temperature. 1: Lequeux et al. 
(1979); 2: French (1980); 3: Kinman \& Davidson (1981); 4: Davidson 
\& Kinman (1985); 5: Dufour et al. (1988); 6, 7, 8: Pagel et al. 
(1992); 9: Skillman \& Kennicutt (1993); 10: Stevenson et al. (1993); 
11: Martin (1996); 12: Legrand et al. (1997). 13: Izotov \& Thuan 
(1998); 14: V\'{\i}lchez \& Iglesias-Paramo (1998). The error bars are 
those quoted by the observers.}
\label{}
\end{figure}

\begin{figure}[htb]        
\centerline{\psfig{figure=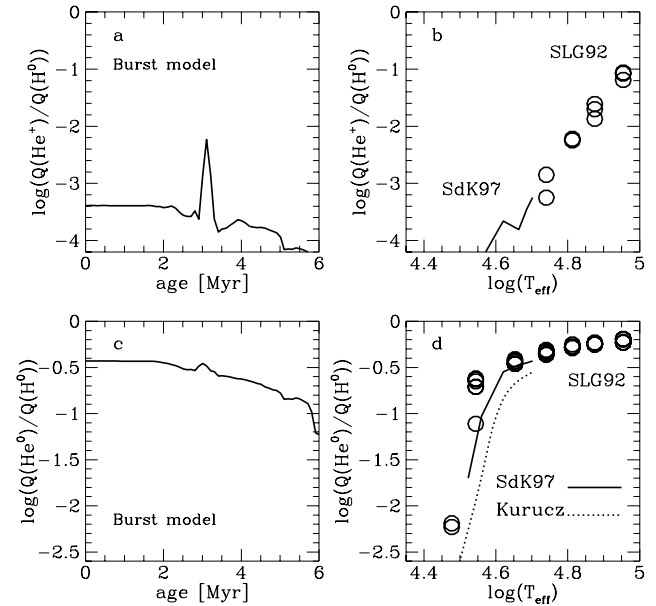,width=8.8cm}}
\caption{ {\bf a)} \qhep/\qh\ for starburst model at Z=0.004 from De Mello
et al.\ (1998) as a function of age. 
{\bf b)} \qhep/\qh\ as a function of effective temperature for
different atmosphere models. SdK97: O dwarf models of Schaerer \& de Koter 
(1997) at Z=0.004, SLG92: WR atmosphere models of Schmutz et al.
\ (1992). Kurucz models have negligible \qhep\ values.
{\bf c)} Hardness parameter \qhe/\qh\ for starburst model as in a.
{\bf d)} \qhep/\qh\ for same atmosphere models as in b and
Kurucz (1991) models for [Fe/H]=-1.5 and $\log g$=5.}
\end{figure}

\begin{figure}[htb]        
\centerline{\psfig{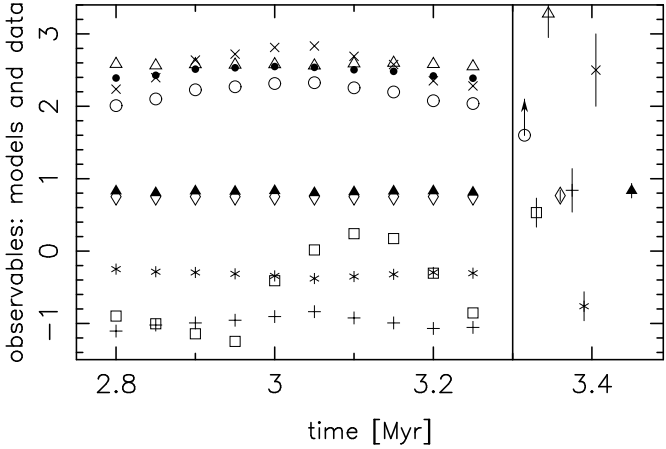}}
\caption{Comparison of observed properties of I~Zw~18 with a series of 
photoionization models of uniform spheres with $n$= 100\cmcub\ and 
a filling factor $\epsilon$=0.01
having, as an ionizing source, the starburst models from de Mello et 
al. (1998) at different ages. Left panel: models. Right panel: 
observations, with indication of the tolerance for the models. 
The following quantities are plotted:
 log F(\Hb) + 15 (open circles), log \Heii/\Hb\ + 2 
(squares), angular radius $\theta$ (crosses), 
\rOiii\ $\times$ 100 (open triangles), \rSii\ (diamonds), 
log (\oiii/\oii)
(black triangles), log (\siii/\sii) (asterisks) and log \oi/\Hb\ +3
(plus)  as a 
function of the starburst age.}
\label{}
\end{figure}

\begin{figure}[htb]        
\centerline{\psfig{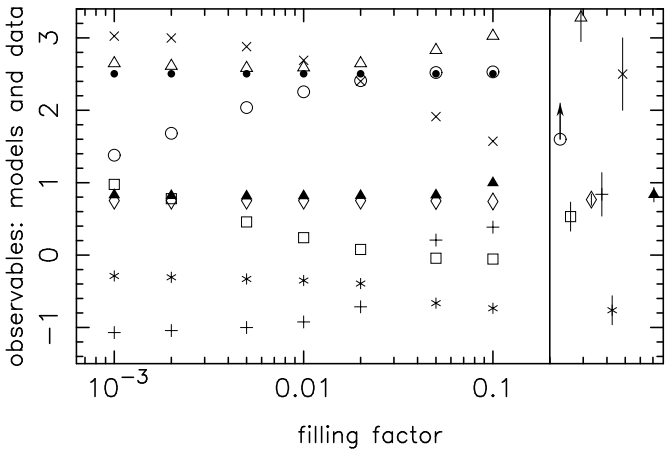}}
\caption{Same as Fig. 4 for a series of models of uniform spheres, with 
$n$= 100\cmcub\ and different filling factors, ionized by the 
radiation field from the model of de Mello et al. (1998)  at an age 
of 3.1 Myr.}
\label{}
\end{figure}

\begin{figure}[htb]        
\centerline{\psfig{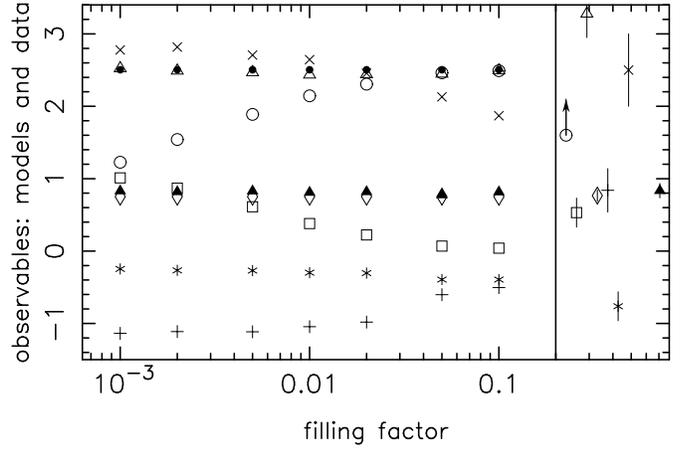}}
\caption{Same as Fig. 5 for a series of models having a shell geometry 
($\theta_{in}$=1.5\arcsec).}
\label{}
\end{figure}

\begin{figure}[htb]        
\centerline{\psfig{figure=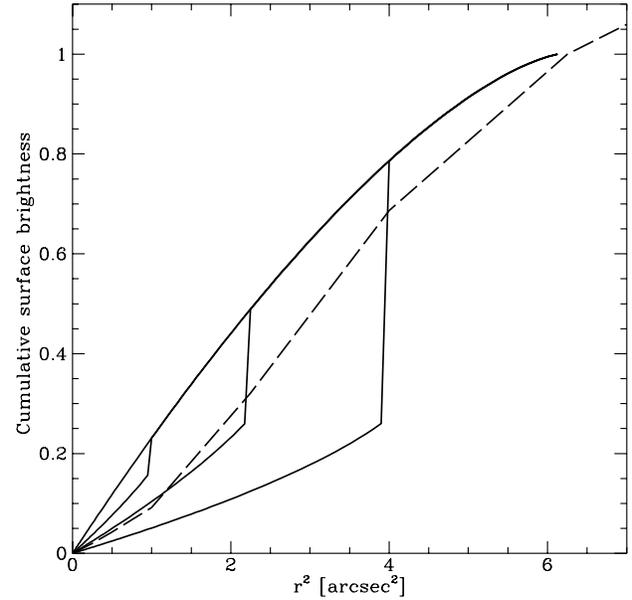,width=8.8cm}}
\caption{Cumulative surface brightness profiles (integrated up to radius $r$)
as a function of $r^2$ in arcsec. All profiles are normalised to $r=2.5$\arcsec.
Solid lines: models for spherical shells of constant density and
various inner radii. Dashed line: observed \Ha\ profile from the {\em HST} image
shown in Fig.\ 1. The observations show a clear deviation from spherical
gas distributions.}
\label{}
\end{figure}

\clearpage

\begin{table*}
\caption{Mean relative ionic abundances and mean ionic temperatures for the best 
uniform sphere model of Fig. 5
}
\begin{flushleft}
\begin{tabular}{lllllllllllll}
\hline
\noalign{\smallskip}  
 & I & II & III & IV & V & VI & VII & VIII & IX \\
\hline
H  & 2.585e-03  &9.974e-01                                                                               \\
   &    16029.  &   17136.                                                                               \\
He & 3.963e-03  &9.658e-01  & 3.023e-02                                                                   \\
   &    16089.  &   16892.  &    24967.                                                                   \\
C  & 7.775e-05  &1.074e-01  & 7.930e-01 &9.525e-02 &4.243e-03                                             \\
   &    15846.  &   16168.  &    16947. &   19230. &   29282.                                             \\
N  & 2.012e-04  &1.047e-01  & 7.823e-01 &1.115e-01 &1.090e-03 &1.863e-04                                  \\
   &    15621.  &   15968.  &    16927. &   19533. &   28682. &   35647.                                  \\
O  & 4.044e-04  &1.105e-01  & 8.750e-01 &1.291e-02 &1.126e-03 &4.627e-05 &6.298e-06                       \\ 
   &    15594.  &   15989.  &    17126. &   26194. &   31082. &   36044. &   40715.                       \\
Ne & 4.870e-05  &4.810e-02  & 9.367e-01 &1.282e-02 &2.286e-03 &8.458e-05 &5.971e-07                       \\
   &    15763.  &   16173.  &    17035. &   25411. &   30494. &   35832. &   41693.                       \\
Mg & 6.431e-04  &6.247e-02  & 8.998e-01 &2.928e-02 &7.267e-03 &5.240e-04 &1.084e-05                       \\
   &    16023.  &   16154.  &    16904. &   23329. &   27900. &   32795. &   38679.                       \\
Si & 9.374e-05  &8.091e-02  & 7.370e-01 &1.374e-01 &4.409e-02 &5.123e-04 &                                \\
   &    16170.  &   16314.  &    16779. &   17814. &   22287. &   30426. &                                \\
S  & 3.276e-06  &2.150e-02  & 5.415e-01 &4.010e-01 &3.550e-02 &4.008e-04 &6.938e-05                       \\
   &    15734.  &   15858.  &    16314. &   17935. &   21174. &   28651. &   37285.                       \\
Cl & 1.374e-05  &4.533e-02  & 5.734e-01 &3.636e-01 &1.670e-02 &9.194e-04 &3.075e-05 &9.525e-06            \\
   &    15835.  &   15938.  &    16396. &   18119. &   23487. &   29260. &   34968. &   40585.            \\
Ar & 2.431e-06  &5.553e-03  & 6.254e-01 &3.578e-01 &9.725e-03 &1.437e-03 &8.322e-05 &3.004e-06 &7.254e-07 \\
   &    15517.  &   15865.  &    16329. &   18257. &   26182. &   29982. &   33949. &   39752. &   43983. \\
Fe & 8.416e-06  &1.722e-03  & 8.484e-02 &8.665e-01 &3.197e-02 &1.204e-02 &2.750e-03 &2.036e-04 &3.611e-06 \\
   &    15512.  &   15518.  &    15860. &   16894. &   22399. &   26289. &   30082. &   34962. &   39424. \\
\noalign{\smallskip}                       
\hline
\end{tabular}
\end{flushleft}
\end{table*}

\begin{table*}
\caption{Mean relative ionic abundances and mean ionic temperatures for the best 
spherical shell model of Fig. 6
}
\begin{flushleft}
\begin{tabular}{lllllllllllll}
\hline
\noalign{\smallskip} 
 & I & II & III & IV & V & VI & VII & VIII & IX \\
\hline
H  & 2.410e-03 & 9.976e-01                                                                               \\
   &    16315. &    16709.                                                                               \\
He & 3.938e-03 & 9.651e-01 & 3.092e-02                                                                    \\
   &    16324. &    16658. &    18322.                                                                    \\
C  & 7.507e-05 & 1.109e-01 & 8.212e-01 & 6.743e-02 & 3.274e-04                                             \\
   &    16188. &    16392. &    16710. &    17188. &    18455.                                             \\
N  & 1.532e-04 & 1.011e-01 & 8.224e-01 & 7.626e-02 & 1.352e-04 & 2.690e-07                                   \\
   &    16013. &    16271. &    16714. &    17216. &    18287. &    18819.                                   \\
O  & 2.960e-04 & 1.073e-01 & 8.868e-01 & 5.670e-03 & 2.764e-05 &                                             \\
   &    15992. &    16277. &    16749. &    18397. &    18731. &                                             \\
Ne & 4.440e-05 & 5.038e-02 & 9.426e-01 & 6.959e-03 & 5.394e-05 & 0. & 0.                        \\
   &    16110. &    16381. &    16714. &    18208. &    18616. &        0. &        0.                        \\
Mg & 6.280e-04 & 6.240e-02 & 9.202e-01 & 1.652e-02 & 2.233e-04 & 4.500e-07 &        0.                        \\
   &    16285. &    16371. &    16705. &    18116. &    18547. &    18831. &        0.                        \\
Si & 1.011e-04 & 7.953e-02 & 7.804e-01 & 1.244e-01 & 1.549e-02 & 1.379e-05 &                                  \\
   &    16431. &    16467. &    16676. &    16956. &    17567. &    18010. &                                  \\
S  & 3.123e-06 & 2.051e-02 & 5.821e-01 & 3.819e-01 & 1.545e-02 & 4.877e-05 & 2.998e-08                        \\
   &    16083. &    16189. &    16466. &    17062. &    17737. &    18475. &    19133.                        \\
Cl & 1.317e-05 & 4.477e-02 & 6.190e-01 & 3.293e-01 & 6.927e-03 & 6.193e-05 & 8.940e-08 &        0.           \\
   &    16195. &    16247. &    16516. &    17102. &    18047. &    18668. &    19015. &        0.           \\
Ar & 1.750e-06 & 5.340e-03 & 6.789e-01 & 3.139e-01 & 1.878e-03 & 2.086e-05 & 3.481e-08 &                     \\
   &    15837. &    16194. &    16499. &    17159. &    18389. &    18745. &    19137. &                     \\
Fe & 5.725e-06 & 1.128e-03 & 7.662e-02 & 8.766e-01 & 4.300e-02 & 2.612e-03 & 3.737e-05 & 2.083e-08 &        0.     \\
   &    15901. &    15917. &    16173. &    16676. &    18213. &    18661. &    18797. &    19168. &        0.   \\
\noalign{\smallskip}                       
\hline
\end{tabular}
\end{flushleft}
\end{table*}

\end{document}